\documentclass[11pt,a4paper]{article}

\usepackage{amsmath}
\usepackage[T1]{fontenc}
\usepackage[utf8]{inputenc}
\usepackage{amsfonts}
\usepackage{amsthm}
\usepackage{amstext}
\usepackage{amssymb}
\usepackage{color}
\usepackage{bbold}
\usepackage{hyperref}
\usepackage{url}

\newcommand{\cc}{\mathbb{C}}
\newcommand{\rr}{\mathbb{R}}

\newcommand{\diag}{\operatorname{diag}}
\newcommand{\rk}{\operatorname{rank}}
\newcommand{\sk}{\operatorname{srank}}
\newcommand{\brk}{\operatorname{\underline{rank}}}

\newtheorem{problem}{Problem}

\newtheorem{theorem}{Theorem}

\newtheorem{lemma}[theorem]{Lemma}
\newtheorem*{lemma*}{Lemma}
\newtheorem{proposition}[theorem]{Proposition}

\newtheorem{remark}[theorem]{Remark}

\newtheorem*{open*}{Open~question}

\DeclareMathOperator{\Ima}{Im}

\begin{document}

\title{On tensor rank and commuting matrices}

\author{Pascal Koiran\footnote{Univ Lyon, EnsL, UCBL, CNRS,  LIP, F-69342, LYON Cedex 07, France. 
Email: {\tt pascal.koiran@ens-lyon.fr}.}}

\maketitle

\begin{abstract}
Obtaining superlinear lower bounds on tensor rank is a major open problem in complexity theory. In this paper we propose a generalization of the approach used by Strassen in the proof of his ${3 \over 2} n$
border rank lower bound.
Our approach revolves around a problem on  commuting matrices:

Given matrices $Z_1,\ldots,Z_p$ of size $n$ and an integer $r>n$, 	 are there commuting { diagonalizable} matrices $Z'_1,\ldots,Z'_p$ 
of size $r$ such that every $Z_k$ is embedded as a submatrix in the top-left corner of $Z'_k$?

As one of our main results, we show that this question always has a positive answer for $r \geq \rk(T)+n$, where $T$ denotes the tensor with slices $Z_1,\ldots,Z_p$. 
Taking the contrapositive, if one can show for some specific matrices $Z_1,\ldots,Z_p$  and a specific integer~$r$ that this question has a negative answer, this yields the lower bound $\rk(T) > r-n$. There is a little bit of slack in the above $\rk(T)+n$ 
bound, but we also provide a number of {\em exact characterizations} of tensor rank and symmetric rank, for ordinary and symmetric tensors, over the fields of real and complex numbers. 
Each of these characterizations points to a corresponding variation on the above approach. 
In order to explain how Strassen's theorem fits within this framework we also provide a self-contained proof of his lower bound.
\end{abstract}

\section{Introduction}

The starting point of this paper is the celebrated lower bound on tensor rank by Strassen~\cite{strassen83}. 
Before giving the statement of his theorem, we recall that a tensor of order 3 and format $m \times n \times p$ can be cut into $p$ "slices." 
Each slice is a $m \times n$ matrix. These are the $z$-slices. One can of course cut $T$ in the two other directions into its $x$-slices 
and $y$-slices.
\begin{theorem}[Strassen] \label{th:strassen}
Let $T$ be a complex tensor of format $n \times n \times 3$  
with $z$-slices denoted  $A_1,A_2, A_3$.
If $A_1$ is invertible,
\begin{equation} \label{eq:strassen}
\rk(T) \geq n + \frac{1}{2} \rk(A_2A_1^{-1}A_3 - A_3A_1^{-1}A_2).
\end{equation}
\end{theorem}
We recall that obtaining rank lower bounds for tensors of order 3 is important because 
this is an adequate complexity measure for 
bilinear computations such as, e.g., matrix multiplication~\cite{Strassen73,BCS}. Moreover, strong enough lower bounds on the rank of (higher order) tensors would yield lower bounds on the size of arithmetic formulas~\cite{raz13}.

Clearly, one cannot hope to obtain more than a   ${3 \over 2} n$ lower bound  by a direct application of Theorem~\ref{th:strassen}. { This maximal lower bound is achieved for instance on the matrix multiplication tensor~\cite{strassen83}.}
Strassen showed that the right-hand side of (\ref{eq:strassen}) is in fact a lower bound on the border rank of $T$. 
This is due to 
the lower-semicontinuity of matrix rank (see \cite[Theorem 4.1]{strassen83} or the appendix  for details).
In the main body of the present paper we will only deal with tensor rank rather than border rank.
Theorem~\ref{th:strassen} proved important for several subsequent results. In particular, it was used by Bläser \cite{Blaser99} in the proof of his ${5 \over 2}n^2$ lower bound on the complexity of matrix multiplication.\footnote{This is still a linear lower bound since the relevant tensor for matrix multiplication is of format $n^2 \times n^2 \times n^2$.}
Generalizations of Theorem \ref{th:strassen} (in its border rank version) 
were given by Landsberg and Manivel~\cite{landsberg08}.
For other approaches to rank lower bounds see e.g.~\cite{alexeev11} or the survey~\cite{blaser14}.

Today, almost 40 years after Strassen's result, we are still unable to prove superlinear lower bounds on tensor rank. In order to achieve this goal, one clearly needs to find a good way of taking into account more than 3 slices (like in Theorem \ref{th:strassen}), and more generally more than a constant number of slices (since the rank of such tensors is linearly bounded).
{ Results in this direction were obtained in a series of papers by Landsberg and his coauthors~\cite{landsberg08,landsberg13,landsberg15} (see~\cite{landsberg17} for the state of the art as of 2017).
 For instance, the "Koszul flattenings" of~\cite{landsberg15,landsbergGCT} yield lower bounds as high as  $(2p+1)n/(p+1)$ for tensors with $2p+1$ slices.
When $p >1$ this is a clear improvement on Theorem~\ref{th:strassen}, but the resulting lower bounds remain linear.
As we will explain in Section~\ref{sec:barriers}, there is a good reason for this state of affairs.}

In this paper we give several characterizations of tensor rank which point to a plausible approach to 
the challenging task of improving on this state of the art, and ultimately of obtaining  superlinear lower bounds.
This approach revolves around variations on the following problem. 
\begin{problem} \label{pb:commute}
Given matrices $Z_1,\ldots,Z_p$ of size $n$ and an integer $r>n$, 	 are there commuting matrices $Z'_1,\ldots,Z'_p$ 
of size $r$ such that $Z_k$ is a submatrix of $Z'_k$ for $k=1,\ldots,p$?
\end{problem}
Here and in the remainder of this paper, when we say that $Z_k$ is a submatrix of $Z'_k$ we mean that $Z_k$ appears in the top-left corner of $Z'_k$, i.e., $(Z'_k)_{ij} = (Z_k)_{ij}$ for $1 \leq i,j \leq n$. We will also say that $Z_1,\ldots,Z_k$ are {\em embedded as submatrices} in $Z'_1,\ldots,Z'_k$.
Theorem~\ref{th:strassen} fits naturally within this framework. Indeed, the quantity $\rk(A_2A_1^{-1}A_3 - A_3A_1^{-1}A_2)$ 
in that theorem can be seen as a measure of the lack of commutativity of the matrices $A_1^{-1}A_2$ and $A_1^{-1}A_3$ 
(this measure is equal to 0 iff the matrices commute). 
In fact, as part of his proof of Theorem \ref{th:strassen}, Strassen offered a solution to Problem~\ref{pb:commute} for the case of two matrices (see Lemma~\ref{lem:strassen} in Section~\ref{sec:strassen}).
{ This approach cannot be pushed much further because Problem~\ref{pb:commute} always has a positive solution
for $r=2n$. We can indeed take:
\begin{equation} \label{eq:2n}
Z'_i = \begin{pmatrix}
Z_i & -Z_i\\
Z_i & -Z_i
\end{pmatrix}
\end{equation}
Since $Z'_i Z'_j=0$ for all $i,j$. For this reason, we will focus on a variation of Problem~\ref{pb:commute} with the additional 
requirement that the $Z'_i$ all be diagonalizable. It turns out that for any tuple of matrices, there is  still a positive solution when
$r$ is large enough.}
Indeed, as one of our main results we provide the following bound.
\begin{theorem} \label{th:embedding}
Let $(A_1,\ldots,A_p)$ be a tuple of $p$ arbitrary matrices of $M_n(K)$. These matrices can be embedded as submatrices in a commuting tuple of 
$p$ diagonalizable matrices of size $\rk(T)+n$, where $T$ denotes the tensor with slices $A_1,\ldots,A_p$.
\end{theorem}
In the above theorem and in the remainder of the paper, $K$ denotes a subfield of the field of complex numbers.
Theorem~\ref{th:embedding} points to the following approach toward lower bounds: if we can show for some specific matrices $Z_1,\ldots,Z_p$  and a specific integer~$r$ that Problem~\ref{pb:commute} has a negative answer, we have obtained the lower bound $\rk(T) > r-n$ where $T$ denotes the tensor with slices $A_1,\ldots,A_p$.
For any matrix tuple, the bound $\rk(T)+n$ is optimal up to the additive term $n$ (see Theorem~\ref{th:tight} in Section~\ref{sec:commuteagain}). As a result, proving a superlinear lower bound on tensor rank is { equivalent} 
to proving  a superlinear lower bound on the smallest $r$ for which Problem~\ref{pb:commute} can be solved with
matrices $Z'_k$ that are all diagonalizable.

This paper provides several variations on Theorem \ref{th:embedding} which point to corresponding variations on this "commutation approach."
For instance, as explained in Section~\ref{sec:results},
we can restrict our search to matrices $Z'_k$ that are symmetric if we seek 
lower bounds on the symmetric rank of symmetric tensors.
{ Besides Strassen's work~\cite{strassen83}, we will see that some elements of the commutation approach can already be found in~\cite{griesser86,landsbergGCT}. More information on the history of lower bounds for tensor rank and border rank
can be found in~\cite{landsbergTensors,landsbergGCT,ottaviani07}.}

\subsection{Results and methods} \label{sec:results}

There is a little bit of slack in Theorem \ref{th:embedding}: the smallest integer for which the embedding is possible lies in the interval $[\rk(T), \rk(T)+n]$. In addition to this result, we provide {\em exact characterizations} of tensor rank in various scenarios. For instance, we give at the end of Section~\ref{sec:rortho} the following characterization of symmetric rank for real symmetric tensors. The symmetric rank, denoted $\sk(T)$ in this paper, is a natural complexity measure for symmetric tensors which satisfies $\rk(T) \leq \sk(T) \leq 4\rk(T)$.
See Section~\ref{sec:background} for background on tensor rank and symmetric rank.
\begin{theorem} \label{th:srealintro}
Let $T$ be a real symmetric tensor of order 3 and size $n$. For any integer~$r$, $\sk(T) \leq r$ if and only if there is  a real symmetric tensor $S$ of order 3 and size $r+n$ 
such that:
\begin{itemize}
\item[(i)] The slices of $S$ commute.
\item[(ii)] $T$ is a subtensor of $S$ in the sense that $T_{ijk} = S_{ijk}$ for all $1 \leq i,j,k \leq n$.
\item[(iii)] Any matrix in the subspace spanned by the slices of $S$ is of rank at most~$r$.
\end{itemize}
\end{theorem}
As another example, returning to ordinary tensors we give the following characterization of tensor rank in Section~\ref{subsec:oind}. 
\begin{theorem} \label{th:indintro}
Let $T$ be a tensor 
of format $n \times n \times p$ over $K$.
Assume moreover that the span of the  $z$-slices of $T$ 
 contains an invertible matrix.
For any integer $r \geq n$, $\rk(T) \leq r$ if and only if there is  a  tensor $S \in K^{r \times r \times p}$ with  $z$-slices $Z_1,\ldots,Z_p$ 
such that:
\begin{itemize}
\item[(i)] The span of $Z_1,\ldots,Z_p$ 
contains an invertible matrix.
\item[(ii)]   For any invertible matrix $Z$ in the span of $Z_1,\ldots,Z_p$, the matrices  $Z^{-1}Z_i$ commute and are diagonalizable over $K$.
\item[(iii)] $T$ is a subtensor of $S$ in the sense that $T_{ijk} = S_{ijk}$ for  $1 \leq i,j \leq n$ and $1 \leq k \leq p$.
\end{itemize}
\end{theorem}
Note that the commutativity condition on the matrices $Z^{-1}Z_i$ in the above theorem  is reminiscent of Strassen's Theorem. 
As we will see, Theorem~\ref{th:indintro} is an important step toward the proof of Theorem~\ref{th:embedding}. 
In a nutshell, to prove the latter theorem we manage to apply Theorem \ref{th:indintro} with $Z$ equal to the identity matrix.
{ This result appears as Proposition~5.1.4.1 in~\cite{landsbergGCT} under the additional assumptions: $K=\cc$, $n=p=r$ and $T$ is "concise" in the sense of \cite[Definition~2.1.7.3]{landsbergGCT}.}

A word is in order regarding the proofs of Theorem \ref{th:srealintro}, Theorem \ref{th:indintro} and of the other exact characterizations is this paper. They are all based on "simple characterizations" of tensor decompositions of a restricted
form. For instance, Theorem \ref{th:srealintro} relies on a characterization of symmetric orthogonal decompositions,
which are of the form
\begin{equation} \label{eq:symdecomp}
T = \sum_{i=1}^r v_i^{\otimes 3}
\end{equation}
where the $v_i$ are nonzero pairwise orthogonal vectors. Not all symmetric tensors admit such a decomposition, 
but those that do admit a very simple characterization: their slices must commute~\cite{boralevi17,koiran2019ortho}.
In general, one cannot expect such a simple characterization for tensor rank or symmetric rank since computing these 
quantities  is NP-hard~\cite{hastad90,schaefer16,shitov16}. 
An arbitrary symmetric decomposition is of the form (\ref{eq:symdecomp}), with the orthogonality constraint on the~$v_i$ removed. From an arbitrary family $v_1,\ldots,v_r$ of vectors of $\rr^n$ we can obtain an orthogonal family in $\rr^{r+n}$
by adding $r$ coordinates to each vector (Section~\ref{sec:rortho}, Lemma~\ref{lem:extension}). In this way we obtain an orthogonal
decomposition of a symmetric tensor of size $r+n$, and a way to transform the characterization of symmetric orthogonal 
decompositions into a characterization of arbitrary symmetric decompositions. 
The proofs of our other results follow a similar pattern. In particular, we give in Section \ref{sec:ind} an alternative 
characterization of the symmetric rank of real symmetric tensors. 
The role of orthogonal decompositions is now played  by {\em independent decompositions}, i.e., instead of an orthogonality
requirement on the $v_i$ we only assume that they are linearly independent.\footnote{ Joseph Landsberg (personal communication) has pointed out that Griesser used a similar method
in the proof of his border rank lower bound (\cite{griesser86} and~\cite[Theorem~5.2.2.1]{landsbergGCT}).}
This leads to a total of 8 different scenarios: we consider symmetric and ordinary tensors, over the fields of real and complex numbers, 
and we characterize their ranks using orthogonal or independent decompositions. For independent decompositions we can handle
all subfields of $\cc$ in a single theorem (like in Theorem~\ref{th:indintro}). We therefore need to  present only 6 exact characterizations of rank and symmetric rank. Toward the proof of Theorem~\ref{th:indintro} we need a "simple characterization" of independent decompositions of ordinary tensors.
 We propose such a characterization in Theorem \ref{th:indordi} since we could not find a suitable one in the literature.
Tensor decomposition algorithms based on similar linear independence assumptions have been proposed for instance in~\cite{lathauwer06,lathauwer04}. 
The characterization in Theorem~\ref{th:indordi} seems especially related to Jennrich's algorithm (see~\cite{moitra18} and the references therein).

{\subsection{Connection to barrier results} \label{sec:barriers}

Many lower bounds in algebraic complexity are obtained by estimating the rank of a matrix $L(f)$ associated to a polynomial $f$, or to a tensor. In order to obtain a lower bound for a "hard polynomial" $\hat{f}$ one must show that $\rk L(\hat{f})$ is "large", and 
that  "small" arithmetic circuits can only compute polynomials $f$ for which $\rk L(f)$ is small.
Very often, the map $f \mapsto L(f)$ is linear. In this case we have a {\em rank method} in the sense of~\cite{efremenko18}.
It is shown in that paper that rank methods face severe limitations. For instance, they cannot prove lower bounds larger than
$8n$ on the rank of order 3 tensors.
An extensive list of lower bound methods belonging to this category can be found in~\cite{efremenko18}. 
It includes for instance the methods of partial derivatives~\cite{NW96}  and shifted partial derivatives~(\cite{GKKS14,fournier15lower} among others).
Strassen's theorem, however, is conspicuously absent from their list and indeed, the proofs of Theorem~\ref{th:strassen} in~\cite{strassen83} or in the present paper follow a very different pattern. It is nevertheless possible to prove a certain version
of Theorem~\ref{th:strassen} by a rank method: see Theorem~\ref{th:rank} in the appendix.

In its full generality, the "commutation approach" proposed in the present paper is immune to the barrier result from~\cite{efremenko18} 
for a very simple reason: as explained after the statement of Theorem~\ref{th:embedding}, proving a lower bound 
on tensor rank (by any approach) is equivalent up to a small additive term to proving a lower bound by the commutation approach.
At this stage, the missing ingredient is a good way of giving a negative answer to Problem~\ref{pb:commute} 
or to the similar problems arising from the characterizations of tensor rank and symmetric rank appearing in Sections~3 to~6.
This was arguably the easier part of the proof of Strassen's theorem (see Lemma~\ref{lem:strassen} in Section~\ref{sec:strassen}), but obtaining a superlinear lower bound is
likely to be much harder.

\subsection{Organization of the paper}

We begin in Section~\ref{sec:background} with some background on the decomposition of ordinary and symmetric tensors. 
We also provide an outline of the proof of Strassen's theorem 
so the reader can better appreciate its relations to Problem~\ref{pb:commute} and Theorem~\ref{th:indintro}. 
The main results of the paper appear in Sections~3 to~6. They are organized according to the different scenarios presented in 
Section~\ref{sec:results}:  we consider symmetric and ordinary tensors, 
and we characterize their ranks using orthogonal or independent decompositions.
The four resulting sections are further divided when necessary into a real and a complex case.
In particular, Theorem~\ref{th:srealintro} is proved in Section~\ref{sec:rortho}, Theorem~\ref{th:indintro} is proved in Section~\ref{subsec:oind} and Theorem~\ref{th:embedding} in Section~\ref{sec:commuteagain}.
In the appendix we complete the proof of Theorem~\ref{th:strassen} and of its border rank version. We also point out some differences with Strassen's original proof, and we give a variation on Theorem~\ref{th:strassen} which can be proved 
by a rank method in the sense of~\cite{efremenko18}. 

\section{Background on tensor rank} \label{sec:background}

Recall that a tensor can be viewed as a multidimensional array with entries in some field $K$. 
The main cases of interest for this paper are $K=\rr$ and $K=\cc$. We will only consider tensors of order 3, i.e., elements of $K^{m \times n \times p}$.
We will sometimes work with square tensors, for which $m=n=p$.
We denote by $M_{n,p}(K)$ the set of matrices with $n$ rows, $p$ columns and entries in~$K$. 
We denote by $M_n(K)$ the set of square matrices of size $n$, by~$GL_n(K)$ the group of invertible matrices of size $n$,
and by $I_n$ the identity matrix of size $n$.

Given 3 vectors $u \in K^m, v \in K^n, w \in K^p$ we recall that their tensor product $u \otimes v \otimes w$ is the tensor of  format $m \times n \times p$
with entries: $T_{ijk}=u_i  v_j w_k$.
By definition, a tensor of this form with $u,v,w \neq 0$ is said to be of {\em rank one}.
The rank of an arbitrary tensor $T$ is defined as the smallest integer $r$ such that $T$ can be written as a sum of $r$ tensors of rank one (and the rank of $T=0$ is 0). 
An elementary counting of the number of independent parameters in such a decomposition shows that most of the square tensors of size $n$ must be of rank $\Omega(n^2)$.  For $K = \cc$, the value of the generic rank 
is known exactly: it is equal to 5 for $n=3$~\cite{strassen83} and to $\lceil n^3/(3n-2) \rceil$ for $n \neq 3$~\cite{lickteig85}. The same paper also gives a simple formula
for the dimension of the set of tensors of rank at most~$r$.
For tensors of order 4 or more, the exact value of the generic rank is not known in general. The existing  results are summarized in~\cite[section~5.5]{landsbergTensors}.

By cutting a tensor $T \in K^{m \times n \times p}$ into its $p$ slices in the $z$ direction we obtain $m \times n$ matrices (the so-called "$z$-slices"). This will allow us to study the properties of $T$ with tools from linear algebra. It is therefore  important
to know what the slices of a tensor look like given a decomposition 
\begin{equation} \label{eq:decomp} 
T=\sum_{i=1}^r u_i \otimes v_i \otimes w_i
\end{equation}
as a sum of rank-1 tensors.
First, we note that the $k$-th $z$-slice of $ u_i \otimes v_i \otimes w_i$ is the rank-one matrix $w_{ik} (u_i v_i^T)$.
As a result, the $k$-th $z$-slice of $T$ is
$$Z_k=\sum_{i=1}^r w_{ik} (u_i v_i^T).$$
This can be written in more compact notation:
\begin{equation} \label{eq:slices}
Z_k  = U^T D_k V
\end{equation}
where $U$ is the matrix with the $u_i$ as row vectors, $V$ is the matrix with the $v_i$ as row vectors and $D_k$ is the diagonal matrix 
$\diag(w_{1k},\ldots,w_{rk})$. We record this observation in the following proposition.
\begin{proposition} \label{prop:slices}
For a tensor $T$ of format $m \times n \times p$ we have $\rk(T) \leq r$ iff there are diagonal matrices $D_1,\ldots,D_p$ of size $r$ and two matrices $U \in M_{r,m}(K)$, $V \in M_{r,n}(K)$ such that the $z$-slices of $T$ satisfy $Z_k  = U^T D_k V$ for $k=1,\ldots,p$.
 In this case, we have a decomposition of $T$ as in~(\ref{eq:decomp}) where the $u_i$ are the rows of $U$, the $v_i$ are the rows of $V$ and  
$D_k = \diag(w_{1k},\ldots,w_{rk})$, i.e., the $k$-th coordinate of~$w_i$ is the $i$-th diagonal entry of $D_k$.
\end{proposition}

A  tensor can be naturally interpreted as the array of coefficients of the trilinear form in $m+n+p$ variables: 
\begin{equation} \label{eq:multilinear}
t(x_1,\ldots,x_m,y_1,\ldots,y_n,z_1,\ldots,z_p)=\sum_{i,j,k} T_{ijk}x_iy_jz_k.
\end{equation}
For a decomposition of $T$ as in~(\ref{eq:decomp}) we have for the corresponding  trilinear form the decomposition:
\begin{equation} \label{eq:setmulti}
t(x,y,z)=\sum_{i=1}^r (u_i^Tx)(v_i^Ty) (w_i^Tz).
\end{equation}
Such an expression is sometimes called a "set-multilinear depth-3 homogeneous arithmetic circuit"~\cite{NW96,raz13}.

\subsection{Symmetric tensors}

A square tensor $T$ is said to be symmetric if $T_{ijk}$ is invariant under all 6 permutations of the indices $i,j,k$.
A symmetric tensor can be interpreted as the array of coefficients of the homogeneous polynomial 
$$t'(x_1,\ldots,x_n)=\sum_{i,j,k=1}^n T_{ijk}x_ix_jx_k.$$
The relation with the multilinear form in (\ref{eq:multilinear}) is that $t'(x)=t(x,x,x)$.
For symmetric tensors one may consider arbitrary decompositions as in (\ref{eq:decomp}), but it is very natural to look for
 {\em symmetric decompositions } where $u_i=v_i=w_i$. Such a decomposition provides a decomposition
 $t'(x)=\sum_{i=1}^r (u_i^T x)^3$ as a sum of cubes of linear forms.
 The {\em symmetric rank}, denoted $\sk(T)$ in this paper,  is the smallest number of terms $r$ in any symmetric decomposition of~$T$.
 By definition we have $\sk(T) \geq \rk(T)$ for any symmetric tensor~$T$ (it was shown only recently that this inequality can be strict~\cite{shitov18}). 
 In the other direction we have $\sk(T) \leq 4\rk(T)$.
 This can be shown by substituting $x=y=z$ in~(\ref{eq:setmulti}) and rewriting each of the $r$ products of linear forms as a sum of 4 cubes using the formula:\footnote{This can be generalized to tensors of any order $d$ using Fischer's formula \cite{fischer94}: we have $\sk(T) \leq 2^{d-1}\rk(T)$.}
 $$24uvw=(u+v+w)^3-(u-v+w)^3-(u+v-w)^3+(u-v-w)^3.$$
 
  In contrast to the ordinary case,  the exact value of the  symmetric rank for a generic symmetric tensor of any size and any order is known thanks to the Alexander-Hirschowitz theorem~\cite{alexander95,brambilla08}. 

\subsection{Outline of the proof of Strassen's theorem} \label{sec:strassen}

In this section we provide an outline of the proof of Theorem~\ref{th:strassen} so the reader can better appreciate its relations to Problem~\ref{pb:commute} and Theorem~\ref{th:indintro}. 
The first step of the proof is a reduction to the case where the first slice $A_1$ is the identity matrix. This is possible because multiplying each slice by $A_1^{-1}$ (or by any invertible matrix) does not change the tensor rank, as can be seen from Proposition~\ref{prop:slices}. Assuming that $A_1=I_n$, it therefore remains to show that: 
$$r=\rk(T) \geq n + \frac{1}{2} \rk(A_2A_3 - A_3A_2).$$
The next step of the proof is akin to Theorem~\ref{th:indintro}: Strassen embeds $T$ in a tensor $S$ of format $r \times r \times 3$
having $I_r$ as its first slice (of course, this would be impossible without the assumption $A_1=I_n$). 
As a result, we can take $Z=Z_1=I_r$ in Theorem~\ref{th:indintro}.(ii) and we conclude that the last two slices of $S$ commute.
More details on this step can be found in the appendix.

The final result therefore follows from the next lemma, which is Strassen's solution to Problem~\ref{pb:commute} for two matrices:
\begin{lemma} \label{lem:strassen}
If $A_2,A_3 \in M_n(K)$ can be embedded as submatrices in two commuting  matrices $A'_2,A'_3 \in M_r(K)$ we must have
$$r \geq n + \frac{1}{2} \rk(A_2A_3 - A_3A_2).$$
\end{lemma}
\begin{proof}
Consider the block decomposition of $A'_2$ and $A'_3$:
$$A'_i=
\begin{pmatrix}
A_i & B_i \\
C_i & D_i
\end{pmatrix}.$$
Since $A'_2$ and $A'_3$ commute we have $A_2A_3-A_3A_2 = B_3C_2-B_2C_3$.
Since $B_2$ and $B_3$ have only $r-n$ columns, $\rk(B_3C_2-B_2C_3) \leq 2(r-n)$.
\end{proof}

{This lemma can prove lower bounds on the size of the embedding up to $r=3n/2$. As explained in the introduction,
for $r=2n$ an embedding in commuting matrices is always possible. For the case of two matrices $A_2,A_3 \in M_n(K)$ considered in Lemma~\ref{lem:strassen}  one may also take:
$$A'_2 =  \begin{pmatrix}
A_2 & A_3\\
A_3 & A_2
\end{pmatrix},\ 
A'_3=  \begin{pmatrix}
A_3 & A_2\\
A_2 & A_3
\end{pmatrix},$$
as an alternative to~(\ref{eq:2n}).}

\section{Symmetric tensors: from orthogonal decompositions 
to arbitrary decompositions} \label{sec:ortho}

In this section we give a proof of Theorem \ref{th:srealintro} and of its complex counterpart (Theorem~\ref{th:scomplexfinal}). As explained in Section~\ref{sec:results}, these results rely on the existence of "simple characterizations" for orthogonal tensor decompositions.

\subsection{Real tensors} \label{sec:rortho}

Following~\cite{boralevi17}, a real symmetric tensor of order $d$ and size $n$ is said to be {\em symmetrically odeco}
if one can write
\begin{equation} \label{eq:sodeco}
T = \sum_{i=1}^k \alpha_i v_i^{\otimes d}
\end{equation}
where $\alpha_i = \pm 1$ and
$v_1,\ldots,v_k$ are nonzero, pairwise orthogonal vectors in~$\rr^n$. 
One may clearly take $\alpha_i=1$ for all $i$ when $d$ is an odd number.
\begin{theorem} \label{th:sodeco}
A real symmetric tensor of order 3 is symmetrically odeco if and only if its slices pairwise commute.
\end{theorem}
This result was first established in \cite{boralevi17} in a different language: instead of commuting slices, they give a characterization based on the associativity of a bilinear map associated to the tensor. The equivalent formulation in Theorem \ref{th:sodeco} is from \cite{koiran2019ortho}.

\begin{theorem} \label{th:sreal}
Let $T$ be a real symmetric tensor of order 3 and size $n$. For any integer $r$, $\sk(T) \leq r$ if and only if there is  a real symmetric tensor $S$ of order 3 and size $r+n$ 
such that:
\begin{itemize}
\item[(i)] The slices of $S$ commute.
\item[(ii)] $T$ is a subtensor of $S$ in the sense that $T_{ijk} = S_{ijk}$ for all $1 \leq i,j,k \leq n$.
\item[(iii)] $\sk(S) \leq r$.
\end{itemize}
\end{theorem}
The following lemma is needed for the proof of Theorem \ref{th:sreal}.
\begin{lemma} \label{lem:extension}
Let $u_1,\ldots,u_r$ be an arbitrary family of vectors in $\rr^n$. There exists a family $v_1,\ldots,v_r$ of pairwise orthogonal vectors of 
$\rr^{r+n}$ such that $u_i$ is the orthogonal projection of $v_i$ on its first $n$ coordinates, i.e., $u_i=(v_{i1},\ldots,v_{in})$. 
\end{lemma}
\begin{proof}
We are looking for a family $w_1,\ldots,w_r$ of vectors of $\rr^{r}$ such that $\langle u_i,u_j \rangle + \langle w_i,w_j \rangle = 0$ for $i \neq j$ since we can then use the coordinates of $w_i$ as the last $r$ coordinates of $v_i$. Let $A \in M_r(\rr)$ be the symmetric matrix 
with entries $A_{ij}=-\langle u_i,u_j \rangle$ . 
Fix a large enough constant $\lambda$ such that $B=A+\lambda I_r$ is positive semidefinite. This matrix must admit a Cholesky decomposition $B=LL^T$ and we can take the rows of $L$ as our vectors $w_i$.
\end{proof}
\begin{proof}[Proof of Theorem \ref{th:sreal}]
Suppose that properties (ii) and (iii) hold for $S$. 
By~(iii), there are vectors $v_1,\ldots,v_r \in {\rr}^{n+r}$ such that $S=\sum_{i=1}^r v_i^{\otimes 3}$.
By~(ii), $\sk(T) \leq r$ since $T=\sum_{i=1}^r u_i^{\otimes 3}$ where $u_i$ is obtained from the first $n$ coordinates of $v_i$:
$u_i=(v_{i1},\ldots,v_{in})$.  Note that this argument does not use~(i).

For the converse, assume that $\sk(T) \leq r$, and let $$T=\sum_{i=1}^r u_i^{\otimes 3}$$ be a corresponding decomposition. Let $S=\sum_{i=1}^r v_i^{\otimes 3}$ where the orthogonal vectors $v_i$ are given by Lemma \ref{lem:extension}. Condition (ii) holds by construction of the $v_i$, and~(i) holds by Theorem \ref{th:sodeco}. Condition (iii) holds as well since $S$ is defined as a sum of $r$ tensors of rank at most 1.
\end{proof}

\begin{remark}
By Theorem \ref{th:sodeco}, a tensor $S$ satisfying only property (i) of Theorem \ref{th:sreal} will already satisfy $\sk(S)  \leq n+r$.

\end{remark}
The characterization in Theorem \ref{th:sreal} is not completely satisfactory because condition (iii) still appeals to the notion of symmetric rank. We would like instead a characterization in terms of properties of the slices only. For this we need two  additional facts.
\begin{lemma} \label{lem:rankr}
Let $K$ be an arbitrary field. Let $(u_1,\ldots,u_r)$ and $(v_1,\ldots, v_r)$ be two families of  vectors of $K^n$, respectively of rank $r_u$ and $r_v$. 
For the  matrix $M=\sum_{i=1}^r  u_i v_i^T$ we have $\rk M \leq \min(r_u,r_v)$. Moreover, if $r_u=r_v=r$ then $\rk M = r$ as well.
\end{lemma}
\begin{proof}
In the expression for $M$ we can rewrite each $u_i$ as a linear combination of the elements of a basis $e_1,\ldots,e_{r_u}$. 
This yields an expression for $M$ as a sum of $r_u$ matrices of rank at most 1, namely, 
$$M=\sum_{i=1}^{r_u}  e_i w_i^T$$
where the $w_i$ are linear combinations of the $v_i$.
 Hence $\rk M \leq r_u$, and $\rk M \leq r_v$ by a similar argument.

Assume now that $r_u=r_v=r$. It remains to show that $\rk M = r$. This is equivalent to $\dim \ker M = n-r$. 
A vector $x \in K^n$ is in the kernel
 if and only if $\sum_{i=1}^{r} (v_i^T x) u_i =0$. Since the $u_i$ are linearly independent, this is equivalent
 to $v_i^T x =0$ for all $i$. Using now the linear independence of the $v_i$, it follows that the solution space is of dimension $n-r$ as needed.
\end{proof}

For application in sections \ref{sec:oortho} and \ref{sec:oind}, the next proposition is stated for ordinary tensors.
In sections \ref{sec:ortho} and \ref{sec:ind} we only need to apply it to symmetric decompositions, namely, to the case $n=p$ and $u_i=v_i=w_i$.
\begin{proposition} \label{prop:span}
Let $K$ be an arbitrary field and let  $(u_1,\ldots,u_r)$, $(v_1,\ldots, v_r)$ be 2 families of  vectors of $K^n$, respectively of rank $r_u$ and $r_v$. Let  $(w_1,\ldots,w_r)$ be a family  of vectors of $K^p$ of rank $r_w$.
From these families, let us  construct the tensor $T=\sum_{i=1}^r u_i  \otimes v_i \otimes w_i$.
The rank of any matrix in the subspace spanned by the $z$-slices of $T$ is at most equal to $\min(r_u,r_v)$. Moreover,  if $r_u=r_v=r_w=r$ then 
there is a matrix of rank $r$ in this subspace.
\end{proposition}
\begin{proof}
Let $Z_1,\ldots,Z_p$ be the $z$-slices of $T$. 
We have already pointed out before (\ref{eq:slices}) that the $k$-th $z$-slice of $ u_i \otimes v_i \otimes w_i$ is $w_{ik} (u_i v_i^T)$. 
As a result any matrix $M$ in the span of $Z_1,\ldots,Z_p$ is a linear combination 
of the matrices $u_i v_i^T$.  It follows from  Lemma \ref{lem:rankr} or directly from~(\ref{eq:slices})  that $\rk M \leq \min(r_u,r_v)$. 
 
Assume now that $r_u=r_v=r_w$
and consider a linear combination of slices 
$$M=\sum_{k=1}^p x_k Z_k = \sum_{i=1}^r \sum_{k=1}^p w_{ik} x_k (u_i v_i^T).$$
Note that the coefficients of the matrices $u_i v_i^T$ in this expression are the entries of the matrix-vector product $W.x$, where $W$ is the matrix with the $w_i$ as rows. Since $W$ is of full row rank, one can obtain any desired values for these~$r$ coefficients by an appropriate choice of $x$. 
If we choose $r$ nonzero values for these coefficients, it follows from Lemma \ref{lem:rankr} that the corresponding~$M$ is of rank $r$.
\end{proof}

As our final characterization of symmetric rank for real tensors, we can now give a proof of Theorem \ref{th:srealintro}. 
For the reader's convenience we reproduce its statement below.
\begin{theorem} \label{th:srealfinal}
Let $T$ be a real symmetric tensor of order 3 and size $n$. For any integer~$r$, $\sk(T) \leq r$ if and only if there is  a real symmetric tensor $S$ of order 3 and size $r+n$ 
such that:
\begin{itemize}
\item[(i)] The slices of $S$ commute.
\item[(ii)] $T$ is a subtensor of $S$ in the sense that $T_{ijk} = S_{ijk}$ for all $1 \leq i,j,k \leq n$.
\item[(iii)] Any matrix in the subspace spanned by the slices of $S$ is of rank at most~$r$.
\end{itemize}
\end{theorem}
\begin{proof}
Suppose that $\sk(T) \leq r$. Properties (i) and (ii) hold for the tensor~$S$ constructed in Theorem~\ref{th:sreal}. Property (iii) follows from the corresponding property of Theorem \ref{th:sreal} and from the first part of Proposition \ref{prop:span}. 

For the converse, assume that properties (i) to (iii) hold  for some tensor~$S$. We just need to show that $\sk(S) \leq r$ since $\sk(T) \leq r$ will then
follow from Theorem \ref{th:sreal}. By Theorem \ref{th:sodeco} and (i), we have a decomposition $S=\sum_{i=1}^k v_i^{\otimes 3}$ where the
$v_i$ are nonzero orthogonal vectors. By the second part of Proposition \ref{prop:span}, there is a matrix of rank $k$ in the span of the slices of~$S$.
This implies $k \leq r$ by (iii), and we have shown that $\sk(S) \leq r$ as needed.
\end{proof}

\subsection{Complex tensors} \label{sec:csymortho}

Symmetric orthogonal decompositions over~$\cc$ were first studied in \cite{koiran2019ortho} in the language of polynomials, where they are called {\em orthogonal Waring decompostions}. The corresponding set of tensors (or polynomials) was denoted $OW_n(\cc)$.
By contrast, \cite{boralevi17} studied {\em unitary} rather than orthogonal decompositions of complex tensors.
Recall from  \cite{koiran2019ortho} that a symmetric tensor $T$ of size $n$ is said to be in $OW_n(K)$ 
if one can write:
\begin{equation} \label{eq:symortho}
T = \sum_{i=1}^n \alpha_i (A e_i)^{\otimes 3}
\end{equation}
where $\alpha_1,\ldots,\alpha_n \in K$, $(e_1,\ldots,e_n)$ is the standard basis of $K^n$ and $A$ is an orthogonal matrix (i.e., $A^TA=I_n$).
For $K=\rr$, we recover the notion of a symetrically odeco tensor from Section \ref{sec:rortho}. In the remainder of this section we will be interested in the case $K=\cc$.
The following characterization is in the style of (\ref{eq:sodeco}) and can be found in \cite[Section 3.2]{koiran2019ortho}.
\begin{proposition} \label{prop:csodeco}
A symmetric tensor $T$ of size $n$ is in $OW_n(\cc)$ if and only if we have
$$T = \sum_{i=1}^k v_i^{\otimes 3}$$
for some integer $k \leq n$, where $v_1,\ldots,v_k$ are pairwise orthogonal non-isotropic vectors of $\cc^n$.
\end{proposition}
Here, orthogonality is defined with respect to the bilinear form $\langle u,v \rangle = u^Tv$ like in the real case. We recall that a vector $u$ is {\em isotropic} if it is self-orthogonal, i.e., $\langle u,u \rangle =0$.
In terms of slices we have the following characterization, also from \cite[Section 3.2]{koiran2019ortho}.
\begin{theorem} \label{th:csodeco}
A  symmetric tensor of size $n$ is in $OW_n(\cc)$ 
if and only if its slices are diagonalizable and pairwise commute.
\end{theorem}
The diagonalizability condition appears in the above theorem because, in contrast to real symmetric matrices, complex symmetric matrices are not always diagonalizable.
The following lemma is the complex analogue of Lemma~\ref{lem:extension}.
\begin{lemma} \label{lem:cextension}
Let $u_1,\ldots,u_r$ be an arbitrary family of vectors in $\cc^n$. There exists a family $v_1,\ldots,v_r$ of pairwise orthogonal non isotropic vectors of 
$\cc^{r+n}$ such that $u_i$ is the orthogonal projection of $v_i$ on its first $n$ coordinates, i.e., $u_i=(v_{i1},\ldots,v_{in})$. 
\end{lemma}
\begin{proof}
We are now looking for a family $w_1,\ldots,w_r$ of vectors of $\cc^{r}$ such that $\langle u_i,u_j \rangle + \langle w_i,w_j \rangle$ is equal to 0 for $i \neq j$, and is different from 0 for $i = j$.
Let $A \in M_r(\cc)$ be the symmetric matrix 
with entries $A_{ij}=-\langle u_i,u_j \rangle$, and let $B=A+I_r$. 
Like any complex symmetric matrix, $B$ admits a decomposition of the form  $B=LL^T$ \cite[Corollary 2.6.6]{horn13},
and we can take the rows of $L$ as our vectors $w_i$ (in particular, we obtain $\langle v_i,v_i \rangle =1$ for all $i$).
\end{proof}
With this lemma in hand, we can now give the complex analogue of Theorem~\ref{th:sreal}.
\begin{theorem} \label{th:scomplex}
Let $T$ be a complex symmetric tensor of  size $n$. For any integer $r$, $\sk(T) \leq r$ if and only if there is  a complex symmetric tensor $S$ of  size $r+n$ 
such that:
\begin{itemize}
\item[(i)] The slices of $S$ are diagonalizable and pairwise commute.
\item[(ii)] $T$ is a subtensor of $S$ in the sense that $T_{ijk} = S_{ijk}$ for all $1 \leq i,j,k \leq n$.
\item[(iii)] $\sk(S) \leq r$.
\end{itemize}
\end{theorem}

\begin{proof}
One shows like in the real case that  $\sk(T) \leq r$ follows from (ii) and (iii).
For the converse, assume that $\sk(T) \leq r$, and let $$T=\sum_{i=1}^r u_i^{\otimes 3}$$ be a corresponding decomposition. Let $S=\sum_{i=1}^r v_i^{\otimes 3}$ where the pairwise orthogonal non-isotropic vectors $v_i$ are given by Lemma \ref{lem:cextension}. Condition (ii) holds by construction of the $v_i$, and~(i) holds by Proposition \ref{prop:csodeco} and Theorem \ref{th:csodeco}.
Condition (iii) holds as well since $S$ is defined as a sum of $r$ tensors of rank 1.
\end{proof}
We conclude Section \ref{sec:csymortho} with an analogue of Theorem \ref{th:srealfinal}.
\begin{theorem} \label{th:scomplexfinal}
Let $T$ be a complex symmetric tensor of  size $n$. For any integer $r$, $\sk(T) \leq r$ if and only if there is  a complex symmetric tensor $S$ of  size $r+n$ 
such that:
\begin{itemize}
\item[(i)] The slices of $S$ are diagonalizable and pairwise commute.
\item[(ii)] $T$ is a subtensor of $S$ in the sense that $T_{ijk} = S_{ijk}$ for all $1 \leq i,j,k \leq n$.
\item[(iii)] Any matrix in the subspace spanned by the slices of $S$ is of rank at most~$r$.
\end{itemize}
\end{theorem}
\begin{proof}
Suppose that $\sk(T) \leq r$. Properties (i) and (ii) hold for the tensor~$S$ constructed in Theorem~\ref{th:scomplex}. Property (iii) follows from the corresponding property of Theorem \ref{th:scomplex} and from the first part of Proposition \ref{prop:span}. 

For the converse, assume that properties (i) to (iii) hold  for some tensor~$S$. We just need to show that $\sk(S) \leq r$ since $\sk(T) \leq r$ will then
follow from Theorem \ref{th:scomplex}. By Theorem \ref{th:csodeco} and (i), we have a decomposition $S=\sum_{i=1}^k v_i^{\otimes 3}$ where the
$v_i$ are pairwise orthogonal non-isotropic vectors. In particular, the $v_i$ must be linearly independent.
By the second part of Proposition \ref{prop:span}, there is a matrix of rank $k$ in the span of the slices of~$S$.
This implies $k \leq r$ by (iii), and we have shown that $\sk(S) \leq r$ as needed.
\end{proof}

\section{Symmetric tensors: from independent decompositions 
to arbitrary decompositions} \label{sec:ind}

The characterizations of symmetric rank in Section~\ref{sec:ortho} relied on earlier results about orthogonal decompositions.
In this section we give an alternative characterization which relies instead on {\em independent decompositions}.
These are symmetric decompositions involving vectors that are assumed to be linearly independent rather than
orthogonal.
The following theorem summarizes some results from~\cite{koiran19derand} regarding these decompositions.
\begin{theorem} \label{th:symequiv}
Let $K$ be the field of real or complex numbers, and let $S$ be a symmetric tensor of size $r$ over $K$ with slices $Z_1,\ldots,Z_r$. There are linearly independent vectors $v_1,\ldots,v_r \in K^r$ such that $S=\sum_{i=1}^r v_i^{\otimes 3}$ if and only if the following two conditions are satisfied:
\begin{itemize}
\item[(i)] The span of $Z_1,\ldots,Z_r$ contains an invertible matrix.
\item[(ii)]   For any invertible matrix $Z$ in this span, the $r$ matrices  ${Z}^{-1}Z_i$ commute and are diagonalizable over $K$.
\end{itemize}
\end{theorem}
\begin{proof}
This is essentially the content of \cite[Theorem 23]{koiran19derand} except that in that reference, property (ii) is replaced by:
\begin{itemize}
\item[(ii')] The slices $Z_1,\ldots,Z_r$ are simultaneously diagonalizable by congruence.
\end{itemize}
Assume first that $K=\cc$.
If (i) and (ii) hold then (ii') must hold by \cite[Theorem 7]{koiran19derand}.  Conversely, if (i) and (ii') hold then  \cite[Theorem 7]{koiran19derand} shows that:
\begin{enumerate}
\item The matrices $Z^{-1}Z_i$ commute for any invertible matrix $Z$ in the span of the $Z_i$.
\item There exists an invertible matrix $Z$ in the span such the matrices $Z^{-1}Z_i$ are diagonalizable.
\end{enumerate}
It is easily checked that  the proof of \cite[Theorem 7]{koiran19derand} implies that the above diagonalizability property holds not only for {\em some} invertible matrix $Z$ in the span, but for any invertible matrix $Z$ in the span. Hence (ii) follows from (i) and (ii'). For the field of real numbers, the arguments are very similar and can be found in \cite[Section 2.3] {koiran19derand}.
\end{proof}
As a variation on Theorem \ref{th:symequiv}, one could replace (i) and (ii) by:
\begin{itemize}
\item[(iii)]   There exists an invertible matrix $Z$ in the space of $Z_1,\ldots,Z_r$ such that that the matrices  ${Z}^{-1}Z_i$ commute and are diagonalizable over $K$.
\end{itemize}
With Theorem \ref{th:symequiv} in hand, we can now give a characterization of symmetric  tensor rank under the relatively mild assumption that the span of slices contains an invertible matrix.
\begin{theorem} \label{th:symind}
Let $K$ be the field of real or complex numbers, and let $T$ be a symmetric tensor of  size $n$ over $K$. 
Assume moreover that the span of the slices of $T$ contains an invertible matrix.
For any integer $r \geq n$, $\sk(T) \leq r$ if and only if there is  a symmetric tensor $S$ of size~$r$ with slices $Z_1,\ldots,Z_r$ such that:
\begin{itemize}
\item[(i)] The span of $Z_1,\ldots,Z_r$ contains an invertible matrix.
\item[(ii)]   For any invertible matrix $Z$ in this span, the matrices  $Z^{-1}Z_i$ commute and are diagonalizable over $K$.
\item[(iii)] $T$ is a subtensor of $S$ in the sense that $T_{ijk} = S_{ijk}$ for all $1 \leq i,j,k \leq n$.
\end{itemize}
\end{theorem}
\begin{proof}
Suppose that $T=\sum_{i=1}^r u_i^{\otimes 3}$ where $u_1,\ldots,u_r \in K^n$. Since the span of the slices of $T$ contains an invertible matrix, the family $u_1,\ldots,u_r$ 
must be of rank $n$ by Proposition~\ref{prop:span}. Consider the $r \times n$ matrix having the $u_i$ as row vectors. We can attach to this matrix
$r-n$ column vectors in order to obtain a matrix $V$ of rank $r$. Let $v_1,\ldots,v_r \in K^r$ be the row vectors of $V$ 
and let $S=\sum_{i=1}^r v_i^{\otimes 3}$. Property (iii) holds for $S$ by construction. Properties~(i) and (ii) follow from Theorem \ref{th:symequiv} since the family $v_1,\ldots,v_r$ is of rank $r$. Note that  (i)  also follows from Proposition \ref{prop:span}.

For the converse, assume that properties (i) to (iii) hold for $S$. By Theorem~\ref{th:symequiv} there are vectors $v_1,\ldots,v_r$ such that 
 $S=\sum_{i=1}^r v_i^{\otimes 3}$. By (iii) we have $T=\sum_{i=1}^r u_i^{\otimes 3}$ 
 where $u_1,\ldots,u_r$ are obtained from the first $n$ coordinates of $v_1,\ldots,v_r$. 
\end{proof}

\section{Ordinary tensors: from orthogonal decompositions to arbitrary decompositions}
\label{sec:oortho}

The characterizations of tensor rank given in this section rely on appropriate generalizations of the orthogonal decompositions
of Section~\ref{sec:ortho} to the setting of ordinary tensors. 

\subsection{Real tensors} \label{sec:roortho}

Following~\cite{boralevi17}, a real ordinary tensor of size $n$ is said to be {\em odeco}
if one can write
\begin{equation} \label{eq:odeco}
T = \sum_{i=1}^ k u_i \otimes v_i \otimes w_i
\end{equation}
where each of the 3 lists $(u_1,\ldots,u_k)$, $(v_1,\ldots,v_k)$, $(w_1,\ldots,w_k)$ is made of~$k$ nonzero, pairwise orthogonal vectors of $\rr^n$.
A characterization of odeco tensors by degree 2 equations was obtained in ~\cite{boralevi17}.
Like in Section \ref{sec:rortho} we will work instead with a characterization in terms of slices:

\begin{theorem} \label{th:odeco}
For a real tensor $T$ of size $n$, the two following properties are equivalent:
\begin{itemize}
\item[(i)] If $(T_1,\ldots,T_n)$ denotes the tuple of $x$-slices of $T$, or its tuple of $y$-slices, or its tuple of $z$-slices then the matrices $T_k T_l^T$ and $T_k^T T_l$ are symmetric for all $k, l \in \{1,\ldots,n\}$.
\item[(ii)] $T$ is odeco.
\end{itemize}
\end{theorem}
This result from \cite[Section 4]{koiran2019ortho} is the ordinary analogue of Theorem \ref{th:sodeco}.
As pointed out in that paper, property (i) implies in  particular that the matrices $T_k T_l^T$ pairwise commute, and 
that the matrices $T_k^T T_l$ pairwise commute.
The following characterization of tensor rank is the ordinary analogue of Theorem \ref{th:sreal}. 
\begin{theorem} \label{th:real}
Let $T$ be a real tensor of size $n$. For any integer~$r$, $\rk(T) \leq r$ if and only if there is  a real tensor $S$ of size $r+n$ 
such that:
\begin{itemize}
\item[(i)] The $x$, $y$ and $z$-slices of $S$ satisfy property (i) from Theorem \ref{th:odeco}.
\item[(ii)] $T$ is a subtensor of $S$ in the sense that $T_{ijk} = S_{ijk}$ for all $1 \leq i,j,k \leq n$.
\item[(iii)] $\rk(S) \leq r$.
\end{itemize}
\end{theorem}
The proof is omitted since it is a straightforward adaptation of the proof of Theorem \ref{th:sreal} (instead of Theorem \ref{th:sodeco}, we just need to evoke Theorem \ref{th:odeco}).
The final result of Section \ref{sec:roortho} is the ordinary counterpart of Theorem \ref{th:srealfinal}.
Its proof is omitted for the same reason.
\begin{theorem} \label{th:realfinal}
Let $T$ be a real tensor of  size $n$. For any integer~$r$, $\rk(T) \leq r$ if and only if there is  a real tensor $S$ of size $r+n$ 
such that:
\begin{itemize}
\item[(i)] The $x$, $y$ and $z$-slices of $S$ satisfy property (i) from Theorem \ref{th:odeco}.
\item[(ii)] $T$ is a subtensor of $S$ in the sense that $T_{ijk} = S_{ijk}$ for all $1 \leq i,j,k \leq n$.
\item[(iii)] Any matrix in the subspace spanned by the $z$-slices of $S$ is of rank at most~$r$.
\end{itemize}
\end{theorem}

\begin{remark}
Let $T$ be a real tensor of  size $n$, and let $S$ be the tensor constructed from $T$ in the above theorem. Property (iii) is about 
the $z$-slices of $S$, but the $x$ and $y$ slices of $S$ also satisfy the same property.
A similar remark will apply to Theorem \ref{th:complexfinal} in Section \ref{sec:coortho}.
\end{remark}

\subsection{Complex tensors} \label{sec:coortho}

Like in the symmetric case, orthogonal decompositions of complex ordinary tensors were first studied in \cite{koiran2019ortho}.
The corresponding set of tensors (or polynomials) was denoted $OT_n(\cc)$.
Let $K$ be the field of real or complex numbers.
Recall from  \cite{koiran2019ortho} that a  tensor $T$ of  size $n$ is said to be in $OT_n(K)$ 
if one can write:
\begin{equation} \label{eq:symortho}
T = \sum_{i=1}^n \alpha_i (A e_i) \otimes (Be_i) \otimes (Ce_i)
\end{equation}
where $\alpha_1,\ldots,\alpha_n \in K$, $(e_1,\ldots,e_n)$ is the standard basis of $K^n$ and $A,B,C$ are three orthogonal matrices.
For $K=\rr$, we recover the notion of an odeco tensor from Section \ref{sec:roortho}. In the remainder of this section we will be interested in the case $K=\cc$.
The following characterization is the ordinary analogue of Proposition \ref{prop:csodeco}.
\begin{proposition} \label{prop:codeco}
A tensor $T$ of size $n$  
is in $OT_n(\cc)$ if and only if we have
$$T = \sum_{i=1}^k u_i \otimes v_i \otimes w_i$$
for some integer $k \leq n$, where each of the 3 lists $(u_1,\ldots,u_k)$, $(v_1,\ldots,v_k)$, $(w_1,\ldots,w_k)$ is made of $k$  pairwise orthogonal non-isotropic vectors of $\cc^n$.
\end{proposition}
In terms of slices we have the following characterization~\cite[Section 4.2]{koiran2019ortho}.
\begin{theorem} \label{th:ot}
  A complex tensor $T$ of  size $n$ admits an orthogonal decomposition iff the $x$-slices of $T$ satisfy the following conditions:
  \begin{itemize}
  \item[(i)] for each $k$, $X_k^T X_k$ is diagonalizable and
    $\rk X_k = \rk X_k^T X_k$,
  \item[(ii)] the matrices $X_k X_l^T$ and $X_k^T X_l$
  are symmetric for all $k,l \in \{1,\ldots,n\}$,
  \end{itemize}
  and the $y$ and $z$-slices satisfy the same conditions.
  \end{theorem}
  The next result is the ordinary analogue of Theorem~\ref{th:scomplex}, and the complex analogue of Theorem~\ref{th:real}.
  \begin{theorem} \label{th:complex}
Let $T$ be a complex tensor of  size $n$. For any integer $r$, $\rk(T) \leq r$ if and only if there is  a complex  tensor $S$ of size $r+n$ 
such that:
\begin{itemize}
\item[(i)] The $x$, $y$ and $z$-slices of $S$ satisfy the conditions of Theorem \ref{th:ot}.
\item[(ii)] $T$ is a subtensor of $S$ in the sense that $T_{ijk} = S_{ijk}$ for all $1 \leq i,j,k \leq n$.
\item[(iii)] $\rk(S) \leq r$.
\end{itemize}
\end{theorem}
The proof is omitted since it is a straightforward adaptation of the proof of Theorem \ref{th:scomplex} (instead of Proposition \ref{prop:csodeco} and Theorem \ref{th:csodeco}, we just need to apply Proposition \ref{prop:codeco} and Theorem \ref{th:ot}).
The final result of Section \ref{sec:coortho} is the ordinary counterpart of Theorem \ref{th:scomplexfinal}, and the complex counterpart of 
Theorem \ref{th:realfinal}. Its proof is omitted for the same reason.
 \begin{theorem} \label{th:complexfinal}
Let $T$ be a complex tensor of  size $n$. For any integer $r$, $\rk(T) \leq r$ if and only if there is  a complex  tensor $S$ of  size $r+n$ 
such that:
\begin{itemize}
\item[(i)] The $x$, $y$ and $z$-slices of $S$ satisfy the conditions of Theorem \ref{th:ot}.
\item[(ii)] $T$ is a subtensor of $S$ in the sense that $T_{ijk} = S_{ijk}$ for all $1 \leq i,j,k \leq n$.
\item[(iii)] Any matrix in the subspace spanned by the $z$-slices of $S$ is of rank at most~$r$.
\end{itemize}
\end{theorem}

\section{Ordinary tensors: from independent decompositions to arbitrary decompositions}
\label{sec:oind}

In this section 
we generalize the notion of {\em independent decomposition} from Section~\ref{sec:ind} to the setting of ordinary tensors.
As it turns out, we only need to assume that each of the two families $(u_1,\ldots,u_r)$,
$(v_1,\ldots,v_r)$ in~(\ref{eq:decomp})  is made of linearly independent vectors; no assumption is made about the 
family $(w_1,\ldots,w_r)$. 
 Then we give a characterization of the set of decomposable tensors, and we use it to prove Theorems \ref{th:embedding} and \ref{th:indintro}.

\subsection{Independent decompositions of ordinary tensors}

\begin{theorem}[simultaneous diagonalization by equivalence] \label{th:simdiag}
  Let $A_1,\ldots,A_k$ be 
  matrices of size $n$
  and assume that their span contains an invertible matrix~$A$.
  There are   diagonal matrices $D_i$ and two nonsingular matrices
  $P,Q \in M_n(K)$ such that 
$A_i = P D_i Q$ for all $i=1,\ldots,k$ if and only if the $k$ matrices $A^{-1}A_i$ ($i=1,\ldots,k$) form a commuting family of diagonalizable matrices.
\end{theorem}
\begin{proof}
We will show this for the special case $A=A_1$. The general case follows easily since the tuple $(A_1,\ldots,A_k)$ is simultaneously diagonalizable by equivalence if and only if the same is true of the tuple $(A,A_1,\ldots,A_k)$.
  
  Suppose first that $A_i = P D_i Q$ where $P,Q$ are nonsingular and the $D_i$ diagonal. Since $A_1$ is invertible, the same is true of
  $D_1$ and we have $A_1^{-1}A_i = Q^{-1}D_1^{-1}D_iQ$.
  These matrices are therefore diagonalizable, and they pairwise commute.

  Assume conversely that the $A_1^{-1}A_i$ form a commuting family of diagonalizable matrices. This is well known to be a necessary and sufficient condition
  for simultaneous diagonalization by similarity~\cite[Theorem~1.3.21]{horn13}: there must exist a nonsingular
  matrix $Q$ and diagonal matrices $D_2,\ldots,D_k$ such that
  $A_1^{-1}A_i = Q^{-1}D_iQ$ for $i=2,\ldots,k$. Let $P=A_1Q^{-1}$.
  We have $A_1=PD_1Q$ where $D_1$ is the identity matrix, and for $i \geq 2$
  we have $PD_iQ = A_1Q^{-1}D_iQ = A_1(A_1^{-1}A_i)= A_i.$
\end{proof}

\begin{theorem} \label{th:indordi}
Let $S$ be a tensor 
of format $r \times r \times p$ over $K$, with an invertible matrix~$Z$ in the span of its $z$-slices.
The following properties are equivalent:
\begin{itemize}
\item[(i)] There is a family  $(w_1,\ldots,w_r)$ of vectors of $K^p$ and  two linearly independent families $(u_1,\ldots,u_r)$, 
$(v_1,\ldots,v_r)$   of vectors of $K^r$ such that
$$S=\sum_{i=1}^r u_i \otimes v_i  \otimes w_i.$$
\item[(ii)]  The $p$ matrices  ${Z}^{-1}Z_i$ commute and are diagonalizable over $K$.
\end{itemize}
\end{theorem}
\begin{proof}
By Proposition~\ref{prop:slices}, (i) is equivalent to the existence of two invertible matrices $U$ and $V$ and of diagonal matrices $D_1,\ldots,D_p$ of size $r$
such that the $z$-zlices of $S$ satisfy $Z_i=U^TD_iV$ for $i=1,\ldots,p$. This is in turn equivalent to (ii) by Theorem~\ref{th:simdiag}.
\end{proof}

\subsection{A characterization of tensor rank} \label{subsec:oind}

We can derive from Theorem \ref{th:indordi} a characterization of tensor rank under the relatively mild assumption that the spans of 
$z$-slices 
contains an invertible matrix. This is the ordinary analogue of Theorem \ref{th:symind}. This result already appears as Theorem~\ref{th:indintro} in Section~1. We reproduce its statement below for the reader's convenience.
\begin{theorem} \label{th:ind}
Let $T$ be a tensor 
of format $n \times n \times p$ over $K$.
Assume moreover that the span of the 
 $z$-slices of $T$ 
 contains an invertible matrix.
For any integer $r \geq n$, $\rk(T) \leq r$ if and only if there is  a  tensor $S \in K^{r \times r \times p}$ with  $z$-slices $Z_1,\ldots,Z_p$ 
such that:
\begin{itemize}
\item[(i)] The span of $Z_1,\ldots,Z_p$ 
contains an invertible matrix.
\item[(ii)]   For any invertible matrix $Z$ in the span of $Z_1,\ldots,Z_p$, the matrices  $Z^{-1}Z_i$ commute and are diagonalizable over $K$.
\item[(iii)] $T$ is a subtensor of $S$ in the sense that $T_{ijk} = S_{ijk}$ for  $1 \leq i,j \leq n$ and $1 \leq k \leq p$.
\end{itemize}
\end{theorem}

\begin{proof}
Suppose that $T=\sum_{i=1}^r u_i \otimes v_i \otimes w_i$ with  $u_i, v_i \in K^n$ and $w_i \in K^p$. Since the span of the 
$z$-slices of $T$ 
contains an invertible matrix, it follows from Proposition~\ref{prop:span} that each of the 2 families $(u_1,\ldots,u_r)$ and $(v_1,\ldots,v_r)$ 
must be of rank $n$. Consider the $r \times n$ matrix having the $u_i$ as row vectors. We can attach to this matrix
$r-n$ column vectors in order to obtain a matrix~$U'$ of rank $r$. We can likewise obtain a matrix $V' \in GL_r(K)$ 
from the~$v_i$.
Let $(u'_1,\ldots,u'_r)$ and $(v'_1,\ldots,v'_r)$ 
be the 2 families of vectors of $K^r$ obtained respectively from the rows of $U'$ and $V'$. 
Finally, let  $S=\sum_{i=1}^r u'_i \otimes v'_i \otimes w_i$. Property~(iii) holds for~$S$ by construction. Properties~(i) and (ii) follow from Theorem~\ref{th:indordi} since the 2 families $(u_1,\ldots,u_r)$ and $(v_1,\ldots,v_r)$ 
are made of linearly independent vectors. 

For the converse, assume that properties (i) to (iii) hold for $S$. By Theorem~\ref{th:indordi} there are vectors $u'_i,v'_i,w_i$ such that 
 $S=\sum_{i=1}^r u'_i \otimes v'_i \otimes w_i$. By (iii) we have $T=\sum_{i=1}^r u_i \otimes v_i \otimes w_i$ 
 where $u_i$ and $v_i$ are obtained respectively from the first $n$ coordinates of $u'_i$ and $v'_i$.
\end{proof}

\subsection{Making slices commute again} \label{sec:commuteagain}

Compared to e.g. Theorem~\ref{th:scomplexfinal}, there is an unpleasant complication in Theorem \ref{th:ind}:
the commutation property that we have obtained is not directly for the slices $Z_i$ of $S$, but for matrices of the form $Z^{-1}Z_i$. 
In Theorem \ref{th:commuteagain} we will obtain a simpler commutation property, directly for the slices of $S$.
The price to pay is that we do not obtain an exact characterization of tensor rank like in Theorems \ref{th:scomplexfinal} or
 \ref{th:ind}.
 Toward the proof of Theorem \ref{th:commuteagain} we need the following simple fact, already used in \cite{strassen83}.
\begin{lemma} \label{lem:identity}
Let $U$ and $V$ be two $r \times n$ matrices 
such that $U^TV=I_n$. Then $r \geq n$, and one can add  $r-n$ columns to $U$ and $V$ in order to obtain 
two $r \times r$ matrices which satisfy $U'^TV'=I_r$.
\end{lemma}
\begin{proof}
We have $r \geq n$ since $\rk(U^TV) \leq \max(\rk(U),\rk(V))$.
{ Let us add to $V$ the columns of a $r \times (r-n)$ matrix $B$ such that $\Ima(B) = \ker (U^T)$.
The resulting matrix $V'$ is invertible since $\Ima(V) \cap \ker (U^T) = \{0\}$.
Its inverse is obtained by adding to $U^T$ the rows of a matrix $A^T$ such that $\ker(A^T) = \Ima(V)$.
}
\end{proof}

\begin{theorem} \label{th:commuteagain}
Let $T$ be a tensor of 
format  $n \times n \times p$ over $K$ 
and let $r=\rk(T)$. 
There exists  a  tensor $S$ 
of format~$(r+n) \times (r+n) \times p$ with  $z$-slices $Z_1,\ldots,Z_{p}$ such that:
\begin{itemize}
\item[(i)]   The  $Z_i$ commute and are diagonalizable over $K$.
\item[(ii)] $T$ is a subtensor of $S$ in the sense that $T_{ijk} = S_{ijk}$ for all $1 \leq i,j \leq n$ and $1 \leq k \leq p$.
\end{itemize}
\end{theorem}
Note that this is just a restatement of Theorem~\ref{th:embedding} since we can form a tensor $T$ from any tuple of matrices.
\begin{proof}
This is a variation on the proof of Theorem \ref{th:ind}, with some additional elements coming from the proof of Strassen's theorem~\cite{strassen83}. Very roughly, the two main steps are: to construct a tensor $T'$  of rank at most $r+n$ 
obtained from $T$ 
by addition of a  $(1+p)$-th slice equal to the identity matrix; and then to apply the construction of Theorem \ref{th:ind} to $T'$ in order to obtain
a tensor $S'$ with $p+1$ slices and a last slice $Z_{p+1}$ equal to the identity matrix. Theorem~\ref{th:ind} applied with $Z=Z_{p+1}$ then shows that the slices of $S'$ must commute. We can therefore take for $S$ the tensor made of the first $p$ slices of $S'$.

Let us now describe the construction of $T'$. 
Consider a decomposition $T=\sum_{i=1}^r u_i \otimes v_i  \otimes w_i$.  We will take
$T'=\sum_{i=1}^{r+n} u'_i \otimes v'_i  \otimes w'_i$ where:
\begin{enumerate}
\item For $i \leq r$, $u'_i=u_i$, $v'_i=v_i$ and $w'_i \in K^{p+1}$ is obtained from $w_i$ by addition of a last coordinate 
$w'_{i,p+1}=1$.
\item For $i>r$, $w'_i$ is equal to  $e_{p+1}$, the $(1+p)$-th vector of the standard basis of $K^{p+1}$. It will be important at the end of the proof that we chose $w'_{i,p+1}=1$ for all $i$.
\end{enumerate}
These two conditions ensure that the first $p$ slices of $T'$ are those of $T$. Moreover, in the last $n$ terms 
$\sum_{i>r} u'_i \otimes v'_i \otimes e_{p+1}$ of the decomposition of~$T'$, 
 the factor $\sum_{i>r} u'_i \otimes v'_i$ can be made equal to any matrix of size $n$ by an appropriate choice of the $u'_i$ and $v'_i$. In particular, we can choose these vectors so that $T'$ has the identity matrix as its last slice.

Next we describe the construction of $S'$ from $T'$.
Since $T'$ has a slice of rank $n$ (the last one),  by Proposition~\ref{prop:span} the two families $(u'_1,\ldots,u'_{r+n})$ 
and $(v'_1,\ldots,v'_{r+n})$ must be of rank $n$.
We can now proceed as in the proof of  Theorem \ref{th:ind}. Namely, let $U'$ be the $(r+n) \times n$ matrix having the $u'_i$ as row vectors.
We can attach to this matrix
$r$ column vectors in order to obtain a matrix $U''$ of rank $r+n$. We can likewise obtain a matrix $V'' \in GL_{r+n}(K)$ 
from the $v'_i$.
 By construction, $T'$ is a subtensor of 
 \begin{equation} \label{eq:stensor}
 S'=\sum_{i=1}^{r+n} u''_i \otimes v''_i \otimes w'_i
 \end{equation}
 where the $u''_i$ and $v''_i$ are respectively the rows vectors of $U''$ and $V''$.  
 Moreover, property (ii) in Theorem \ref{th:ind} shows that we have the following property for the slices $Z_i$ of $S'$:
 \begin{itemize}
 \item[(ii')]  For any invertible matrix $Z$ in the span of $Z_1,\ldots,Z_{p+1}$, the matrices  $Z^{-1}Z_i$ commute and are diagonalizable over $K$.
 \end{itemize}
 Therefore, if we can take $Z=I_{r+n}$ in (ii') we have obtained Theorem~\ref{th:commuteagain}.(i). As a result, all that remains to be done is to show that we can obtain $Z_{p+1}=I_{r+n}$. We will obtain this property by a careful choice of the matrices $U''$ and~$V''$.  
 First, note that since $I_n$ is the last slice of $T'$ we have $$U'^TD_{p+1}V'=I_n$$ by~(\ref{eq:slices}), where
 $D_{p+1} = \diag(w_{1,p+1},\ldots,w_{r+n,p+1})$. Remembering that we chose $w_{i,p+1}=1$ for all $i$, this yields $U'^TV'=I_n$. 
 By Lemma~\ref{lem:identity}  we can add $r$ columns to $U'$ and $V'$ in order to obtain $(r+n)\times (r+n)$ matrices $U''$ and $V''$ which satisfy
 $U''^T V'' = I_{r+n}$. We conclude that the last slice of the tensor~$S$ in~(\ref{eq:stensor}) is equal to 
 $U''^TD_{p+1}V''=U''^TV''=I_{r+n}$, as needed. Finally, as announced at the beginning of the proof we take  for $S$ the tensor made of the first $p$ slices of $S'$.
 \end{proof}

As our final result we observe that Theorem~\ref{th:commuteagain} (i.e., Theorem~\ref{th:embedding})  is tight up to the additive term $n$.
\begin{theorem} \label{th:tight}
Let $(A_1,\ldots,A_p)$ be a tuple of $p$ matrices of $M_n(K)$.
If these matrices can be embedded as submatrices in a commuting tuple of 
$p$ diagonalizable matrices of size $N$, we must have $N \geq \rk(T)$ where $T$ denotes the tensor with slices 
$(A_1,\ldots,A_p)$.
\end{theorem}
\begin{proof}
Let $(Z_1,\ldots,Z_p)$ be the commuting tuple and let $S$ be the tensor with slices $Z_1,\ldots,Z_p$. Since the $Z_k$ commute and are diagonalizable, they are simultaneously diagonalizable: there are diagonal matrices $D_1,\ldots,D_p$ and  an invertible matrix $P$ such that $Z_k =  P^{-1} D_k P$ 
for $1 \leq k \leq p$.
By Proposition \ref{prop:slices} we have $\rk(S) \leq N$, and $\rk(T) \leq N$ follows since $T$ is a subtensor of $S$.
\end{proof}

{\small

\section*{Acknowledgments}

Discussions with Joseph Landsberg led to a more extensive presentation of prior work, and in particular of rank methods
and barrier results. The embedding in~(\ref{eq:2n}) was communicated by Jeroen Zuiddam.

%\bibliographystyle{plain}
%\bibliography{../../Biblio/biblio}}
%\newpage

}

\appendix

\section*{Appendix: Completing the proof of Strassen's theorem}

Most of the elements of the proof of Theorem~\ref{th:strassen} can be found scattered in Sections~\ref{sec:background} and~\ref{sec:oind}. In this appendix we tie up the loose ends. 
Recall from Section~\ref{sec:strassen}  that the second step of the proof is a version of Theorem~\ref{th:indintro} for a tensor with the identity matrix as its first slice. The precise statement is as follows; we will need to apply it to a tensor $T$ with only $p=3$ slices.
\begin{proposition} \label{prop:strassen}
Let $T$ be a tensor 
of format $n \times n \times p$ over $K$, with the identity matrix $I_n$ as its first slice.
Assume moreover that $T$ admits a decomposition as a sum of $r$ rank 1 tensors as in~(\ref{eq:decomp}) 
with $w_{i1} \neq 0$ for all $i$. Then there is a tensor $S \in K^{r \times r \times p}$ with  $z$-slices $Z_1,\ldots,Z_p$ 
such that:
\begin{itemize}
\item[(i)] $Z_1=I_r$.
\item[(ii)]   The $Z_i$ commute and are diagonalizable over $K$.
\item[(iii)] $T$ is a subtensor of $S$ in the sense that $T_{ijk} = S_{ijk}$ for  $1 \leq i,j \leq n$ and $1 \leq k \leq p$.
\end{itemize}
\end{proposition}
\begin{proof}
We will assume that $w_{i1}=1$ for all $i$. This is without loss of generality since $w_i$ can be multiplied in~(\ref{eq:decomp}) by $w_{i1}^{-1}$ and $u_i$ (or $v_i$) by $w_{i1}$ if necessary. We can now proceed like in the proof of Theorem~\ref{th:commuteagain}, with the role of the additional $(1+p)$-th slice now played by $Z_1$. 
Namely, let $U,V$ be the two $r \times n$ matrices having respectively the $u_i$ and $v_i$ as row vectors.
Since $T$ has $I_n$ as its first slice we have $U^TV=U^TD_1V=I_n$ by~(\ref{eq:slices}). 
By Lemma~\ref{lem:identity} we can add $r-n$ columns to $U$ and $V$ in order to obtain 
two $r \times r$ matrices which satisfy $U'^TV'=I_r$.
Then we define $S=\sum_{i=1}^r u'_i \otimes v'_i \otimes w_i$ like in the proof of Theorem~\ref{th:ind}, where
the $u'_i, v'_i$ are the row vectors of $U'$ and $V'$. Property~(iii) holds by construction of $S$ and (i) follows from~(\ref{eq:slices}). Property~(ii) follows from Theorem~\ref{th:indordi} applied with $Z=Z_1$.
\end{proof}
As explained in Section~\ref{sec:strassen}, after applying the above proposition we can conclude with Lemma~\ref{lem:strassen}. We therefore have a proof of Theorem~\ref{th:strassen} {\em under the hypothesis that 
 $w_{i1} \neq 0$ for all $i$}. 
  We complete the proof with a perturbation argument. Consider $r$ sequences $(w_i^{(k)})_{k \geq 0}$ of vectors 
 of $\cc^n$ such that $w_i = \lim_{k \rightarrow +\infty} w_i^{(k)}$ and  $w_{i1}^{(k)} \neq 0$ 
 for all $i$ and for all $k$. Our result can be applied to the tensors
 $$T^{(k)} =\sum_{i=1}^r u_i \otimes v_i \otimes w_i^{(k)}:$$
 we have 
  \begin{equation} \label{eq:perturb}
\rk(T^{(k)}) \geq n + \frac{1}{2} \rk(A_2^{(k)}B_1^{(k)}A_3^{(k)} - A_3^{(k)}B_1^{(k)}A_2^{(k)})
\end{equation}
where the $A_i^{(k)}$ are the slices of $T^{(k)}$ and $B_1^{(k)}$ is the inverse of $A_1^{(k)}$. 
Note that the inverse is well-defined for large enough $k$ since the slices of $T^{(k)}$ converge to those of $T$.
Moreover we have 
\begin{equation} \label{eq:lower}
\rk(A_2^{(k)}B_1^{(k)}A_3^{(k)} - A_3^{(k)}B_1^{(k)}A_2^{(k)}) \geq  \rk(A_2A_1^{-1}A_3 - A_3A_1^{-1}A_2)
\end{equation}
 for large enough $k$ by lower semicontinuity of matrix rank, and 
$\rk(T^{(k)}) \leq r = \rk(T)$ by construction of $T^{(k)}$. \qed

 Instead of this pertubation argument,  Strassen used an induction on the number of indices for which $w_{i1}=0$. 
 Another difference with our proof lies in Proposition~\ref{prop:strassen}: 
he did not consider the case $p>3$, which  is not needed
 for the proof of Theorem~\ref{th:strassen}. Also, he did not phrase this argument as an embedding of $T$ 
 in a bigger tensor $S$.
 Instead, he constructed directly the two matrices $Z_2$ and $Z_3$, denoted $\hat{B}$ and $\hat{C}$ in his paper.
 
 \subsection*{Border rank}
 
 Recall that a tensor $T$ is said to be of border rank at most $r$ if there exists a sequence $(T^{(k)})$ of tensors converging to $T$ such that $\rk(T^{(k)}) \leq r$ for all~$k$. The border rank is denoted $\brk(T)$; by definition, $\brk(T) \leq \rk(T)$. 
 For the sake of completeness, we show that the  the right-hand side of (\ref{eq:strassen}) not only provides a lower bound on 
 $\rk(T)$, but also on $\brk(T)$.
 This follows from the argument that we just used to complete the proof of Theorem~\ref{th:strassen}. Consider indeed 
 any sequence   $(T^{(k)})$ of tensors converging to $T$. As pointed out above, the first slice of $T^{(k)}$
 must be invertible for all large enough $k$, and then (\ref{eq:perturb}) holds by Theorem~\ref{th:strassen}. 
 Finally, (\ref{eq:lower}) holds for large enough $k$ by lower semicontinuity of matrix rank.\qed
 
 Instead of this direct argument, Strassen completed the proof of his border rank lower bound by a more algebraic argument.
 He showed that the set of tensors of border rank at most $r$ is included in a variety defined by an explicit system of polynomial equations, obtained from~(\ref{eq:strassen}) by clearing out the denominator $\det(A_1)$ from $A_1^{-1}$ on the right-hand side. 
 These equations are called the "Strassen equations" in~\cite{landsbergTensors,landsberg08}.
 One can consult these two references  for other proofs of Theorem~\ref{th:strassen} in its border rank version.

{ \subsection*{A variant of Strassen's theorem by a rank method}

Finally, we prove a version of Strassen's theorem by a rank method following Ottaviani~\cite{ottaviani07} and Landsberg~\cite{landsbergGCT}.
In order to have a rank method in the sense of~\cite{efremenko18} we need a flattening, i.e., a linear map $L$ from a space $\cal T$ of tensors to a space $\cal M$ of matrices.
If $\rk(L(T)) \leq r_1$ for every rank-one tensor $T \in \cal T$, it follows from sub-additivity of matrix rank that 
$\rk(T) \geq \rk(L(T)) /r_1$ for every $T \in \cal T$. 
For tensors with 3 slices,  one gets an interesting result from a map $L:K^{n \times n \times 3} \rightarrow M_{3n}(K)$
defined on page 32 of~\cite{landsbergGCT}. This map sends a tensor~$T$ with slices $A_1, A_2, A_3$ to the matrix
$$L(T)=\begin{pmatrix}
0 & A_2 & -A_3\\
A_3 & A_1 & 0\\
A_2 & 0 & A_1
\end{pmatrix}.$$
It turns out that the image of  a rank-one tensor by $L$ is always of rank 2 (this is Exercise~2.4.1.1 of \cite{landsbergGCT}).
We therefore obtain the following lower bound.
\begin{theorem} \label{th:rank} 
For a tensor $T$ of format $n \times n \times 3$ we have $\rk(T) \geq \rk(L(T)) / 2$.
\end{theorem}
The connection with Theorem~\ref{th:strassen} follows from equation~(2.4.4) in~\cite{landsbergGCT}:
$$\det L(T) = \det(A_1)^2 \det(A_2A_1^{-1}A_3 - A_3A_1^{-1}A_2).$$
As a result, for a tensor such that 
$\det(A_1) \neq 0$,
Theorem~\ref{th:strassen} provides its maximal lower bound ($3n/2$) if and only if the same is true of Theorem~\ref{th:rank}. 
If $\det(A_2A_1^{-1}A_3 - A_3A_1^{-1}A_2) = 0$, it is not clear how the lower bounds of Theorems~\ref{th:strassen} and~\ref{th:rank} compare.

Theorem~\ref{th:rank} can be generalized to tensors with more than 3 slices (see \cite{landsberg15} and Section~2.4.2 of~\cite{landsbergGCT}).  The resulting  "Koszul flattenings" yield lower bounds as high as  $(2p+1)n/(p+1)$ for tensors with $2p+1$ slices.
}

\end{document}